%% file: H3467_cor.tex
\documentclass[]{aa}

\usepackage{psfig}
\usepackage{rotating}
\usepackage{graphicx}
\usepackage{natbib}
\usepackage[innercaption]{sidecap}

        \def\mum{${\rm \, \mu m}$}                            
        \def\cm3{${\rm cm^{-3}}$}             

        \newcommand{\Tel}{$T_{\rm e}$}                      
        \newcommand{\Teff}{$T_{\rm eff}$}                   
                               
        \newcommand{\nh}{$n_{\rm H}$}                      
        \newcommand{\rgal}{$R_{\rm Gal}$}

        \newcommand{\HII}{\ion{H}{ii}}
        \newcommand{\HeI}{\ion{He}{i}}
        
        \newcommand{\NeII}{[\ion{Ne}{ii}]}
        \newcommand{\NeIII}{[\ion{Ne}{iii}]}

        \newcommand{\ArII}{[\ion{Ar}{ii}]}
        \newcommand{\ArIII}{[\ion{Ar}{iii}]}
        \newcommand{\SIV}{[\ion{S}{iv}]}

        \newcommand{\SIII}{[\ion{S}{iii}]}

        \def\neratio{[\ion{Ne}{iii}]/[\ion{Ne}{ii}]}
        \def\arratio{[\ion{Ar}{iii}]/[\ion{Ar}{ii}]}
        \def\sratio{[\ion{S}{iv}]/[\ion{S}{iii}]}

\begin{document}

\title{The stellar content, metallicity and ionization structure of 
  \HII\ regions
    \thanks{Based on observations with ISO, an ESA project with
      instruments funded by ESA Member States (especially the PI 
      countries: France, Germany, the Netherlands and the United
      Kingdom) and with the participation of ISAS and NASA.} }

        \author{N.L.\,Mart\'{\i}n-Hern\'{a}ndez\inst{1} 
        \and R.\,Vermeij\inst{1}
        \and A.\,G.\,G.\,M.\,Tielens\inst{1,2} 
        \and J.\,M.\,van der Hulst\inst{1}
        \and E.\,Peeters\inst{2,1}
        }

\offprints{N.L.\,Mart\'{\i}n-Hern\'{a}ndez (leticia@astro.rug.nl)}

\institute{ 
  Kapteyn Institute, P.O. Box 800, 9700 AV Groningen, The Netherlands
  \and SRON, National Institute for Space Research, P.O. Box 800,
  9700 AV Groningen, The Netherlands}

\date{Received date; accepted date}

\titlerunning{Stellar content, metallicity and ionization structure of \HII\ regions} 
\authorrunning{N.L.\,Mart\'{\i}n-Hern\'{a}ndez et al.}

\abstract{
Observations of infrared fine-structure lines provide direct information
on the metallicity and ionization structure of \HII\ regions and
indirectly on the hardness of the radiation field ionizing these
nebulae.  We have analyzed a sample of Galactic and Magellanic Cloud \HII\
regions observed by the Infrared Space Observatory (ISO) to examine 
the interplay between stellar content,
metallicity and the ionization structure of \HII\ regions. The observed
\SIV\,10.5/\SIII\,18.7 \mum\ and \NeIII\,15.5/\NeII\,12.8 \mum\ line ratios 
are shown to be highly correlated over
more than two orders of magnitude. We have compared the observed line
ratios to the results of photoionization models using different stellar
energy distributions. The derived characteristics of the ionizing star
depend critically on the adopted stellar model as well as the (stellar)
metallicity. We have compared the stellar effective temperatures
derived from these model studies for a few well-studied \HII\ regions with
published direct spectroscopic determinations of the spectral type of
the ionizing stars. This comparison supports our interpretation that
stellar and nebular metallicity influences the observed infrared ionic 
line ratios. We can
explain the observed increase in degree of ionization, as traced by
the \sratio\ and \neratio\ line ratios, by the hardening
of the radiation field due to the decrease of metallicity. The
implications of our results for the determination of the ages of
starbursts in starburst galaxies are assessed.
\keywords{Stars: atmospheres -- Stars: early-type --
          ISM: abundances -- ISM: \HII\ regions --
          Galaxies: starburst -- 
          Galaxies: individual: Milky Way, Magellanic Clouds}}

\maketitle

\section{Introduction}
\label{section:intro}

The spectral type of the ionizing stars and the stellar and nebular
metallicity are intimately interwoven in controlling the ionization 
structure of
\HII\ regions. Hotter stars will ionize trace elements to higher
ionization stages. Metallicity, on the other hand, modifies the
spectral energy distribution (SED) of the ionizing stars and thus, 
influences the
ionization structure indirectly. In particular, line blocking/blanketing is
directly coupled to the metallicity and controls the stellar wind of
massive stars. The stellar wind, in its turn, also modifies the SED 
of the star. The interlocking of these different
effects hampers the determination of the 
stellar content from ionic line observations.
This has clear repercussions on the study of starburst regions in
galaxies as well, where the stellar content is used as an indicator of 
the starburst age.   
Current models of stellar atmospheres have started to take the effects of 
metallicity into account \citep[e.g.][]{schaerer97,pauldrach01}.
Such models can fit the observed line ratios in \HII\ regions reasonable well
\citep[e.g][]{giveon02,morisset:paperiii}. The derived spectral
types for the ionizing sources, however, depend on the stellar model
adopted and can be very different. 

\input{H3467T1.tex}

The Infrared Space Observatory \citep[ISO,][]{kessler96} has provided a 
powerful tool for
the study of the ionization structure of \HII\ regions
\citep[][hereafter MHP02]
{giveon02,martin:paperii}. 
The ratios of fine-structure
lines originating in the ionized gas (for instance,
\ArIII\,9.0/\ArII\,7.0 \mum, \SIV\,10.5/\SIII\,18.7 \mum\
and \NeIII\,15.5/\NeII\,12.8 \mum) are widely used to constrain 
the properties of the ionizing stars in these nebula 
\citep[e.g.][]{watarai98, takahashi00, morisset:paperiii}.
The systematic studies of the ionization
structure of \HII\ regions via these ionic line ratios 
as a function of Galactic center distance
in the Milky Way by \cite{martin:paperii} 
and in the Large and Small Magellanic Clouds
by \cite{vermeij:analysis} 
have now provided us with a
`natural' sample of \HII\ regions over a wide range in metallicity. Such a
sample allows us to study the effect of stellar type and metallicity
on the ionization structure. Moreover, while the
ionizing stars in most \HII\ regions are heavily extincted in the
visible by tens to hundreds of magnitudes, the opening up of the infrared
window has favoured in recent years the use of infrared spectra to
infer the spectral types of the
ionizing stars. An important effort has now started to identify the
ionizing stars of Galactic \HII\ regions in the K-band  
\citep{hanson96,watson97b}
and the first results are now becoming available \citep{kaper02a,kaper02b}.

In this paper, we examine this interplay between stellar content,
metallicity and ionization structure of \HII\ regions based upon the ISO
sample. This paper is structured as follows. Sect.~\ref{section:data} describes
the combined sample of Galactic and LMC/SMC \HII\ regions; 
Sect.~\ref{section:ionization} shows the correlation between the observed
\neratio\ and \sratio\ line ratios. Sect.~\ref{section:seds} and 
Sect.~\ref{section:metallicity} investigate the influence of the SEDs and 
the stellar/nebular metallicity, respectively, on the ionization structure.
Sect.~\ref{section:starburst} discusses the implications for the
interpretation of the spectra of starburst galaxies.
Finally, Sect.~\ref{section:summary} discusses and summarizes the results.

\section{The sample}
\label{section:data}

We combined the ISO Short Wavelength Spectrometer 
 \citep{deGraauw96} observations of the 
Galactic and Magellanic Cloud \HII\
regions as  described in the catalogues by \cite{peeters:catalogue}
and  \cite{vermeij:data}. These catalogues present the
atomic fine-structure line fluxes of 43 Galactic nebulae  and 12 Magellanic
Cloud sources. The first analysis of the observed  fine-structure
lines is presented in MHP02 and \cite{vermeij:analysis}. 
This analysis  includes,
among others, the derivation of elemental abundances and a
discussion on the ionization state of the nebulae. 

From this combined set of \HII\ regions, we selected those for which both 
information on their elemental abundances (through Ne/H) and ionization state 
(through the \SIV\,10.5/\SIII\,18.7\,\mum\ and 
\NeIII\,15.5/\NeII\,12.8\,\mum\ line ratios) could be
derived. The selected \HII\ regions are listed in
Table~\ref{table:sample}, where the source name, the distance  of
the Galactic objects from the center of the Galaxy and the ISO pointing
coordinates are given.  
We chose to use Ne/H as a tracer of the nebular metallicity. Ne/H can
be determined from the IR fine-structure lines to higher precision
than other elements because it is less sensitive to ionization correction
factors and assumptions regarding electron densities  
(see MHP02 for a complete discussion).  Moreover, as a
primary element, the abundance of neon closely follows that of oxygen
\citep[e.g.][and references therein]{henry99}.
As far as the choice of the ionization tracers is concerned, the choice 
of the \sratio\ and \neratio\ line ratios was an obvious one as \arratio\
has been detected in only a few LMC \HII\ regions.

   \begin{SCfigure*}[1.0][!ht]
   \includegraphics[width=8.5cm]{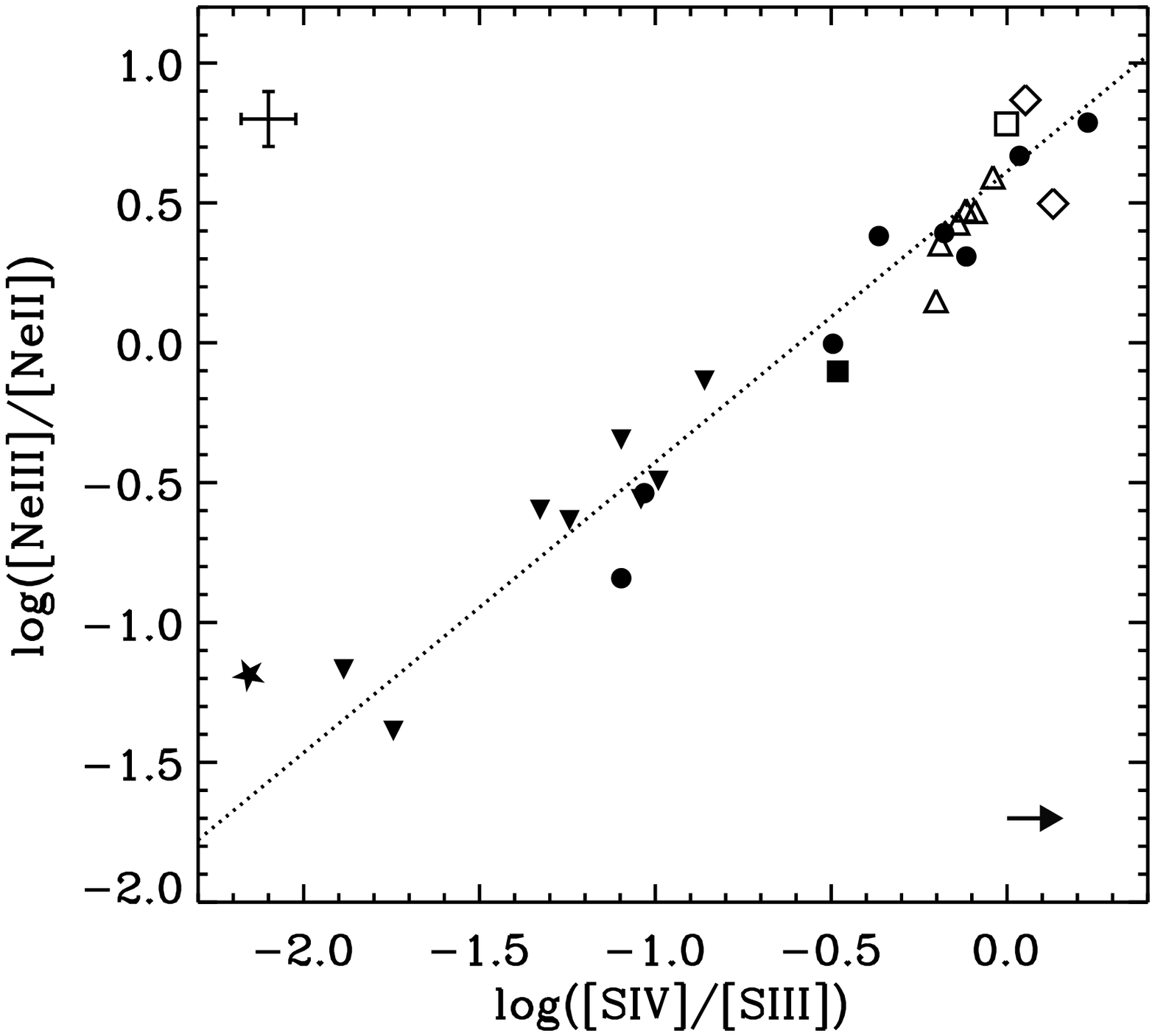}
   \vspace{-0.3cm}
     \caption{Relationship between the  
        \SIV\,10.5/\SIII\,18.7\,\mum\ and \NeIII\,15.5/\NeII\,12.8\,\mum\
        line ratios for the combined sample of \HII\ regions. Indicated by
        various symbols are the Galactic nebulae at \rgal$< 7$ kpc 
        (solid triangles), the Galactic nebulae at \rgal$> 7$ kpc (solid 
        circles), the LMC nebulae (open triangles, except 30\,Doradus,
        which is indicated by an open
        square) and 
        the SMC nebulae (open diamonds). 
        To avoid confusion, only the averaged location of the 4 SWS positions
        in 30\,Doradus is indicated. 
        The positions of Sgr\,A$^*$  
        \citep{lutz96} and the Orion nebula
        \citep{simpson98} are indicated by a solid star and a solid square,
        respectively.
        The dotted line is a least squares fit to the data. 
        A typical error bar of the ISO data is given in the upper
        left corner. The arrow in the lower right corner indicates the
        extinction correction due to a $A_{\rm K}=2$ mag.
        \vspace{1cm}}
      \label{fig:ioniz}
   \end{SCfigure*}

In total, our sample consists of 20 Galactic \HII\ regions located at 
Galactocentric distances between 0 and 15 kpc, 7 \HII\ regions in the 
LMC (including 4 positions in 30\,Doradus) and 2 in the SMC.
Also included are the SWS observations of  
Sgr\,A$^{\star}$, our own Galactic Center \citep{lutz96}, and
the Orion nebula, for which an ``integrated'' infrared 
spectrum was obtained by  \cite{simpson98} using the Midcourse Space Experiment
(MSX) satellite. The MSX field of view (6\arcmin $\times$ 9\arcmin)
included most of the Orion nebula.

\section{Variations in the ionization structure}
\label{section:ionization}

The ionization structure of a nebula depends basically on the shape of the SED
and the nebular geometry. In the case of an ionized sphere of constant
gas density and  filling factor, the geometrical effect can be defined by
the ionization parameter $U=Q_{\rm H}/(4\pi R^2nc)$,
where $Q_{\rm H}$ is the number of stellar photons above 13.6 eV
emitted per second, $R$ is the Str\"omgren radius, $n$ is the gas
density and $c$ is the speed of light. The ionization structure
can be traced by the ratio of successive stages of ionization $X^{+i}$
and $X^{+i+1}$ of a given element. Such a ratio depends, for a given
$U$, on the number of photons able to ionize $X^{+i}$ 
\citep{vilchez88}. 

In Fig.~\ref{fig:ioniz}, we present 
the relation between the \sratio\ and the \neratio\ line ratios for the 
combined sample of \HII\ regions in the Milky Way and 
the LMC/SMC.  
The line ratios have not been corrected for extinction. Because the
lines \NeII\ 12.8 and \NeIII\ 15.5 \mum\ are very close in wavelength,
the \neratio\ is practically insensitive to extinction. However, the
differential extinction between the \SIV\ 10.5 and \SIII\ 18.7 \mum\
lines ($\sim 0.08 A_{\rm K}$ if a standard extinction curve is
considered, see MHP02) can affect the \sratio\ line ratio. The effect of this
differential extinction is indicated in the lower right corner of 
Fig.~\ref{fig:ioniz} for a typical $A_{\rm K}$ of 2 magnitudes (MHP02).  

The data shown in Fig.~\ref{fig:ioniz} span a range in ionization of more than 
2 orders of magnitude. The LMC/SMC points nicely overlap the Galactic trend
at the high ionization end. 
The least squares fit to the data,
displayed as a dotted line, gives a slope 
of $1.04 \pm 0.05$ and an intercept of $0.61 \pm 0.04$.
Fig.~\ref{fig:ioniz} also illustrates the
change in the degree of ionization of the Galactic \HII\ regions
with increasing distance from the Galactic center. The more highly
ionized objects are located at larger Galactocentric distances, while
the less highly ionized objects, including Sgr\,A$^*$, are located 
at smaller distances. 
In line with the
known metallicity gradient in the Galaxy
\citep[e.g. see review by][and references therein]{henry99,rolleston00}, 
the location of the high metallicity Galactic Center at the
low ionization end of the correlation and the low metallicity LMC/SMC
\HII\ regions at the high ionization end suggests that metallicity is
somehow involved in the changes of ionization degree, and hence, 
in the hardening of the radiation field, shown in Fig.~\ref{fig:ioniz}.

We consider in the next sections both the influence of the SED
and the metallicity on the ionization structure of \HII\ regions.

\section{Influence of the SEDs}
\label{section:seds}

   \begin{figure}[!ht]
     \includegraphics[width=8.5cm]{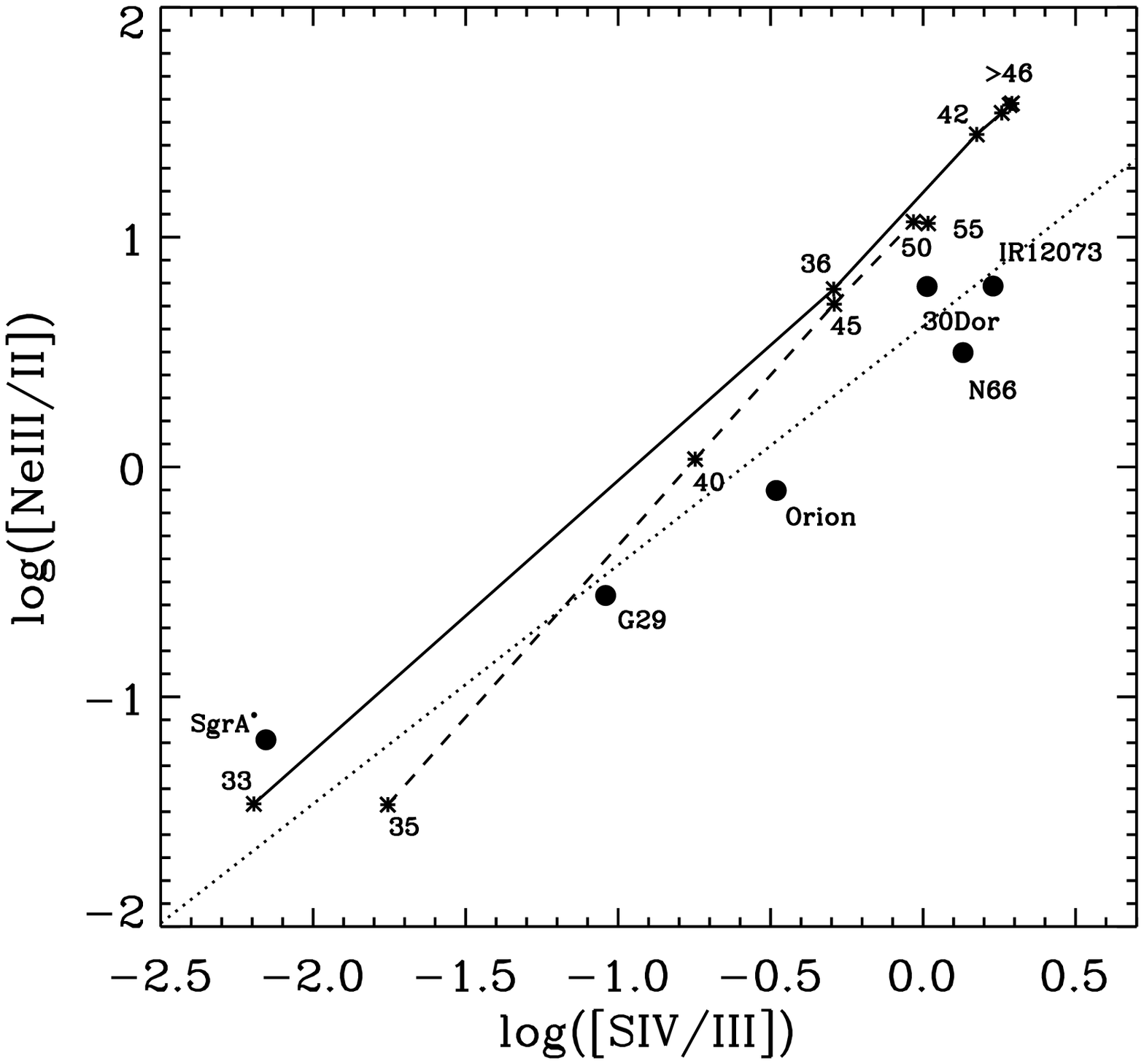}
     \caption{Dependence of the \sratio\ and
       \neratio\ line ratios
       on the stellar \Teff\ (crosses) as given
       by the CoStar (solid lines) and Pauldrach (dashed lines)
       SEDs. Only photoionization models for \nh=$10^3$ \cm3\ are
       plotted.
       The observed line ratios for reference sources
       (Sgr\,A$^*$, G29.96$-$0.02, the Orion nebula, IRAS 12073$-$6233,
       N66 and 30\,Dor) are indicated by solid circles. The dotted line is
       the least squares fit to the data plotted in Fig~\ref{fig:ioniz}.}
     \label{fig:models}
   \end{figure}

In order to investigate the dependence of the \neratio\ and \sratio\
line ratios on the SED, a set of
nebular models has been calculated with the photoionization code
CLOUDY \citep{ferland98} using MICE, the IDL interface for CLOUDY
created by 
H. Spoon\footnote{see http://www.astro.rug.nl/$\sim$spoon/mice.html}. 
The parameters varied in the
modelling are the stellar effective temperature  (\Teff) and the
ionization parameter ($U$). 
We compute nebular models for a static, spherically symmetric, homogeneous  
gas distribution with one ionizing star in the center. An inner cavity  
with a radius equal to 10$^{17}$\,cm is set, while the outer radius of 
the \HII\ region is defined by the position where the electron temperature,
\Tel,
reaches 100\,K. Two different stellar atmosphere models are taken to
describe the SED: the CoStar models by 
\cite{schaerer97} and the models by \cite{pauldrach01}. Previous
studies of these ionization ratios have concentrated on either the
CoStar models \citep[e.g.][MHP02]{morisset:paperiii}) 
or on the Pauldrach models
\citep{giveon02}.
For a  detailed study of the influence of these and other stellar
SEDs on the ionization ratios, we refer
to the work by Morisset et al. (in prep.).
We use stellar models for main sequence (dwarf) stars, i.e. CoStar
models A2, B2, C2, D2, E2 and F2, and Pauldrach models D-35, D-40,
D-45, D-50 and D-55. The metallicity of
the star and the nebula is chosen to be solar.
The number of hydrogen ionizing photons emitted by the central 
source is fixed to the typical value of 
$Q_{\rm H}=10^{48}$\,photons\,s$^{-1}$. The parameter $U$ is varied by 
changing the hydrogen density (\nh). For each of the stellar atmosphere 
models selected,  which span a range in \Teff\ from approximately 30
to 55 kK, photoionization models are calculated for 
\nh$=10^2, 5 \times 10^2, 10^3, 5 \times 10^3$  and $10^4$ \cm3. 

   \begin{figure}[!hb]
     \includegraphics[width=8.7cm]{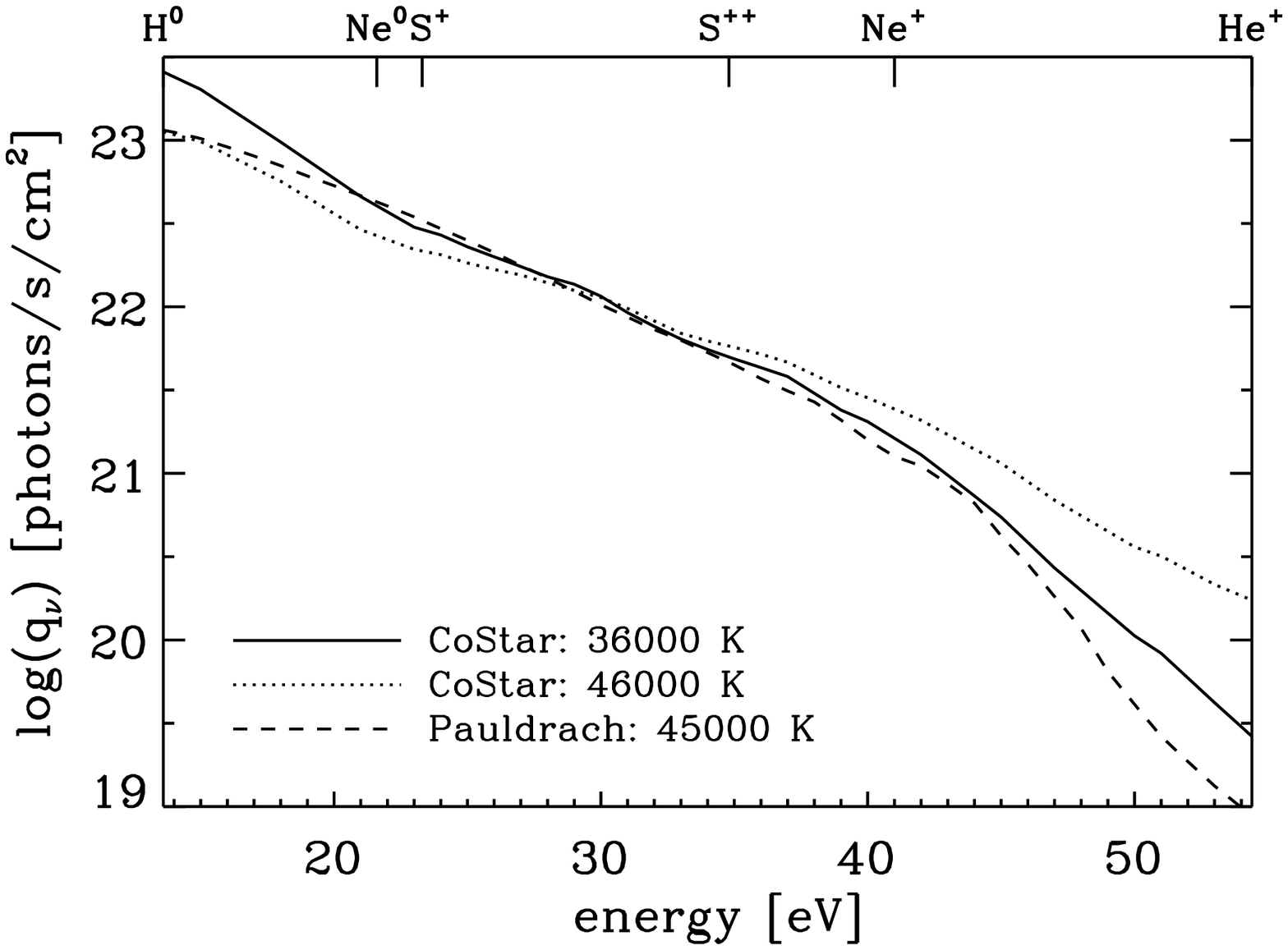}
     \caption{The number of ionizing photons in $\rm\,cm^{-2} s^{-1}$
       above a certain energy from the CoStar B2 (36000 K), D2 (46000 K) 
       and Pauldrach
       D-45 (45000 K) SED models is plotted as a
       function of this energy. The curves are normalized so that the
       total number of hydrogen ionizing photons, $Q_{\rm H}$, is
       $10^{48}$ photons s$^{-1}$, the value used in the photoionization models.
       These curves define, for a given
       ionization parameter $U$, the ionization structure in the
       nebula. Labeled on top of the figure are the ionization
       energies for H$^0$, Ne$^0$, S$^{+}$, S$^{++}$, Ne$^+$ and He$^+$.}
     \label{fig:qv}
   \end{figure}

The output of this grid of models for \nh=10$^3$ \cm3\ is presented in
Fig.~\ref{fig:models}. 
The variation of \neratio\ and \sratio\ with
\Teff\ is shown separately for the two selected SED models.
An increase in density of two orders of magnitude produces an increase of the
line ratios of at most 0.5 dex
approximately along the lines drawn in
Fig.~\ref{fig:models}. 
For reference, the observed line
ratios of the Galactic sources Sgr\,A$^*$, G29.96$-$0.02 (IRAS\,18434$-$0242), 
the Orion nebula and IRAS\,12073$-$6233, the SMC \HII\ region
N66, and 30\,Doradus
are indicated by solid circles. Sgr\,A$^*$ and IRAS\,12073 delimit the observed
parameter space for \neratio\ and \sratio\ as they are, respectively,
the lowest and highest ionized source in Fig.~\ref{fig:ioniz}.

\input{H3467T2.tex}

Several points can be gleaned from inspection of Fig.~\ref{fig:models}: 

{\it (a)} The \neratio\ and \sratio\ line ratios strongly depend on
the stellar \Teff, but they can no longer discriminate between
models when most of the Ne and S is in the higher ionization
stage \citep[see also][MHP02]{giveon02}. When this happens, 
these line ratios remain constant even with
large changes in the SED. When using the CoStar
SEDs, both \neratio\ and \sratio\ become insensitive to the stellar
\Teff\ for \Teff $> 42$ kK, while when using the Pauldrach SEDs, the
hardness of the stellar ionizing flux changes less rapidly with \Teff,
and it is not until \Teff $> 50$ kK that the star fully ionizes
S$^{++}$ and Ne$^{+}$. 

{\it (b)} The slope of the observed log(\neratio) versus
log(\sratio) is practically equal to 1 (cf., Fig~\ref{fig:ioniz}), suggesting
that the spectral hardening affects equally the range of ionizing
energies for \ion{S}{iii} ($h\nu > 35$ eV) and \ion{Ne}{ii} ($h\nu > 41$ eV).  
The model trends are, however, slightly steeper than the
data, i.e. for a given \Teff\ the models overpredict
\neratio\ (underpredict \sratio) by a factor up to 5 for a given
\sratio\ (\neratio). 
As has been discussed
by \cite{oey00} and \cite{morisset:paperiii}, there is an indication that the
CoStar models overpredict the stellar spectrum at these ionizing energies.  

{\it (c)} When using the CoStar SEDs to describe the
emergent flux, practically the entire observed
ionization range spanning more than 2 orders
of magnitude can be modeled
by a rather narrow \Teff\ range between $\sim$ 33 kK and $\sim$ 42
kK. This effect is less dramatic with   
the Pauldrach SEDs, for which the entire range up to 55 kK is
suitable for modelling the observed line ratios.

We see that large differences appear
when we compare the predictions by the CoStar models  to those
by the Pauldrach ones for a given \Teff. These differences are
primarily caused by the different treatment of the line
blocking and blanketing between the two SED models. 
This causes the Pauldrach models
to have a much softer spectrum than the analogous
CoStar star at a similar \Teff.  Note, for instance, the case of the
CoStar SED with \Teff=36000 K (model B2) and the Pauldrach SED with
\Teff=45000 K (model D-45)
in Fig.~\ref{fig:models}, which predict practically the same values
for \sratio\ and \neratio. 

It is instructive to calculate the integrated ionizing photon flux 
above a certain energy as a function of this energy
for these two models (CoStar B2 and Pauldrach D-45). 
The number of ionizing photons in
$\rm cm^{-2}\,s^{-1}$ with energies larger than $h\nu_i$ is
defined as $q_{\nu_i}=\int_{\nu_i}^{\infty} {F_{\nu}/(h\nu) {\rm d\nu}}$, 
with $F_{\nu}$ being the astrophysical flux in 
$\rm erg~s^{-1}~cm^{-2}~Hz^{-1}$.
The results are shown in Fig.~\ref{fig:qv}.
For a significant comparison with the photoionization model results,
the total number of hydrogen ionizing photons has been scaled to $10^{48}$
photons s$^{-1}$, the value used in the models. We note that models
differ in their stellar radii.
The \sratio\ and \neratio\ line ratios are basically sensitive to the
slope of the curve between 
the energies able to produce and ionize S$^{++}$ and Ne$^+$,
 i.e. between 23.3 and 34.8 eV, and 21.6 and 41.0 eV, respectively.
The slopes of the CoStar B2 and the
Pauldrach D-45 SEDs are very similar between these energies; 
hence, their match in the  predicted \sratio\ and
\neratio\ line ratios. 
The ionization structure of a nebula with such values of
\sratio\ and \neratio\ can, therefore, be equally well described by 
any of these
two SEDs. However, the interpretation of this ionization structure in
terms of the  stellar properties, e.g. \Teff, depends on the stellar
model adopted. 
For comparison, we also show the shape of the hotter CoStar model D2,
with a \Teff\ of 46000 K. The ionizing flux of this hotter CoStar model
starts to be higher than the corresponding Pauldrach model above
$\sim$ 30 eV. This hotter CoStar model will produce, consequently, 
a larger amount of S$^{+3}$ and Ne$^{++}$.

   \begin{figure}[!hb]
     \includegraphics[width=8.5cm]{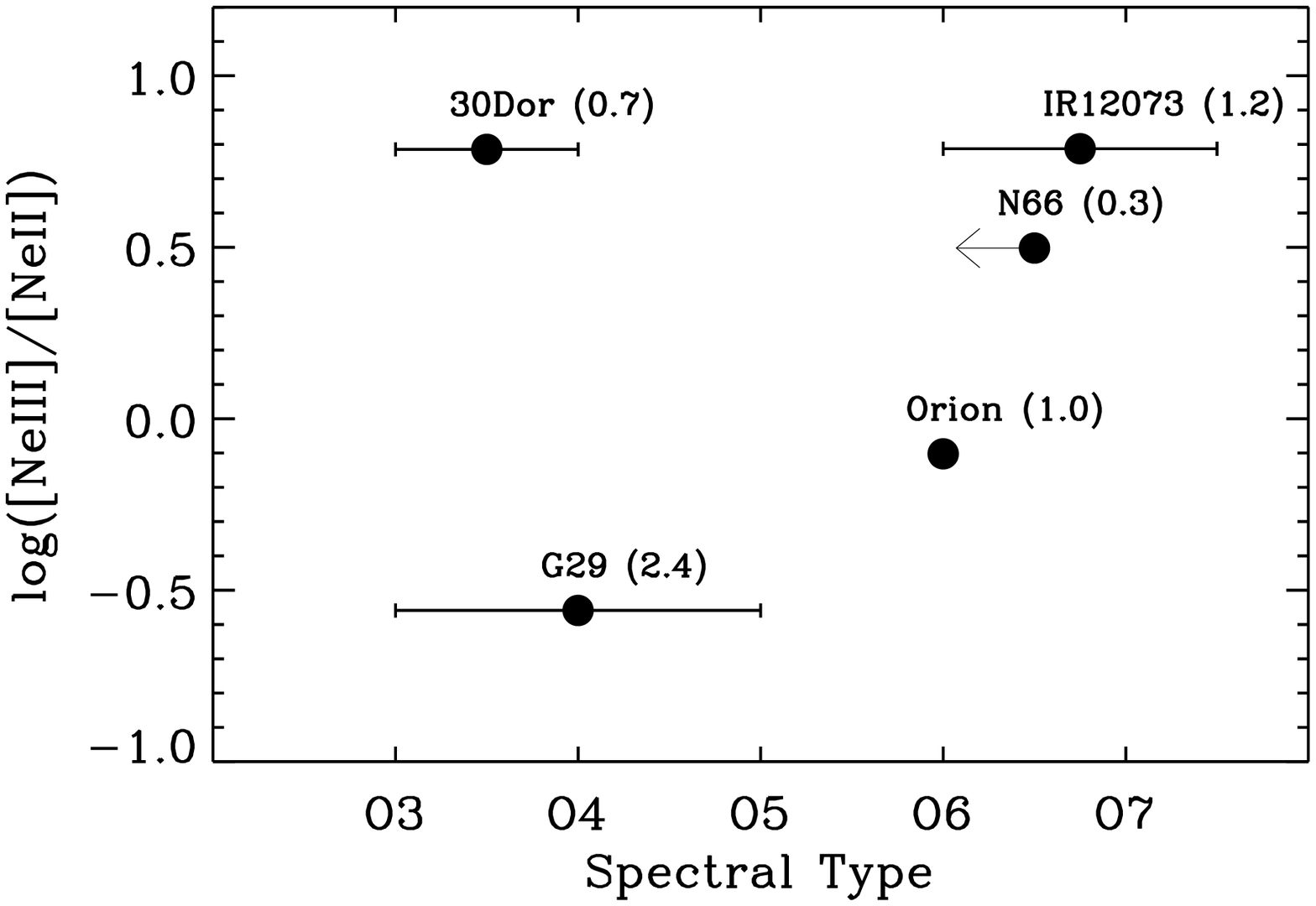}
     \caption{The \neratio\ line ratio
       is compared to the
       observed spectral type for G29.96$-$0.02, the Orion nebula,
       IRAS\,12073$-$6233, N66 and 30\,Doradus. The numbers in
       brackets indicate the Ne/H abundance by number in units
       of $10^{-4}$. References for the Ne/H
       abundances are: G29 and IRAS\,12073 (MHP02); N66 and 30\,Dor
       \citep{vermeij:analysis}; Orion \citep{simpson98}.}
     \label{fig:sptype}
   \end{figure}

   \begin{SCfigure*}[1.0][!ht]
   \includegraphics[width=8.5cm]{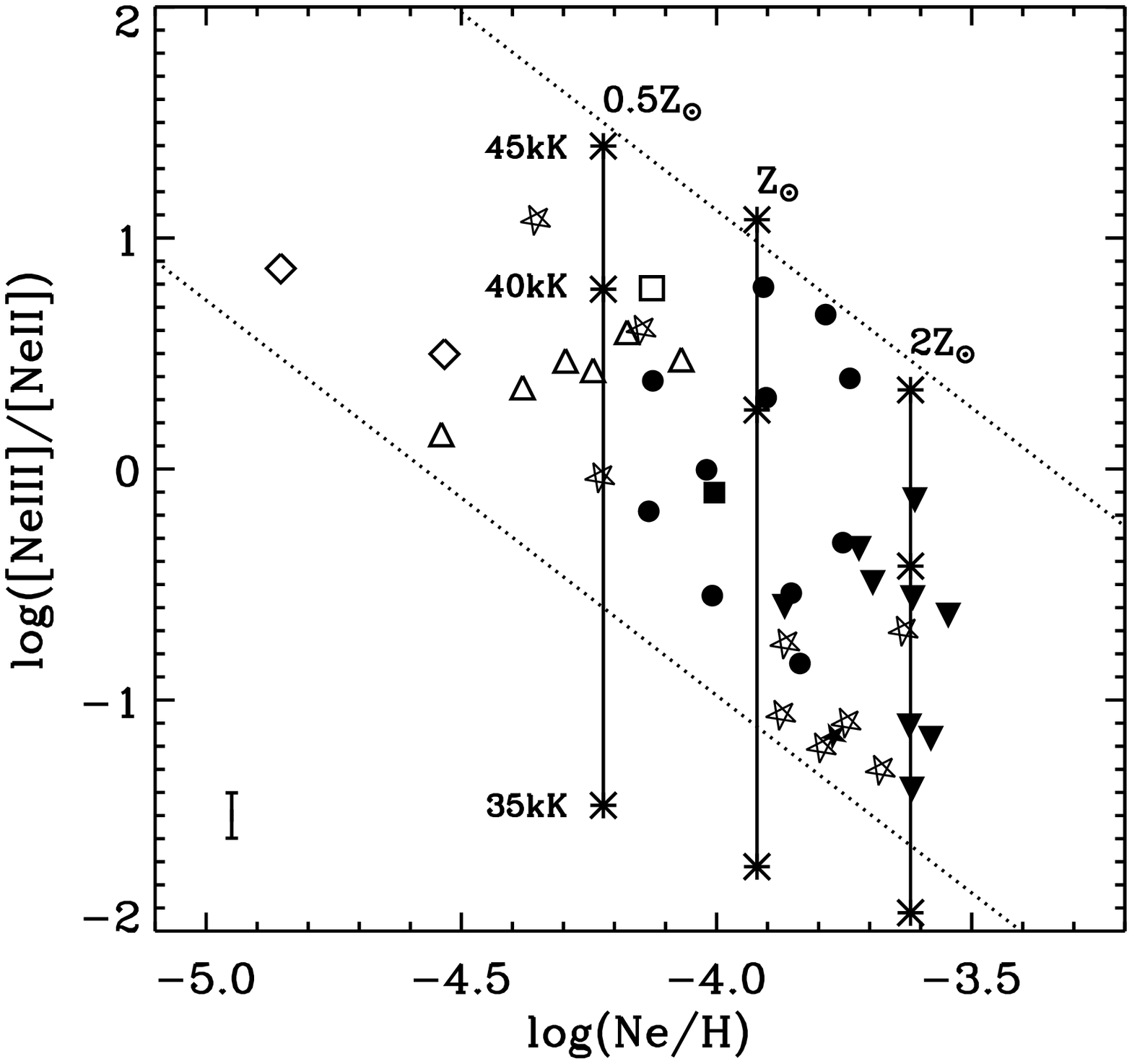}
       \caption{Relationship between the
        \neratio\ line ratio
        and the metallicity indicator Ne/H for the combined sample of
        \HII\  regions. Symbols are as in Fig.~\ref{fig:ioniz}. 
        The data collected on starburst galaxies (open stars) are compared to
        the  Galactic and LMC/SMC \HII\ regions. 
        The parallel
        lines delineate the parameter space span up by the data. 
        A typical error
        bar for log(\neratio) is shown in the left bottom corner. No error bar
        for Ne/H is indicated as the main uncertainty (which can be as
        large as a factor of 2 for the Galactic sources, 
        see Sect.~\ref{section:metallicity}) 
        comes from the abundance derivation method itself and is very hard to
        determine. The solid lines show the results of the
        photoionization models calculated by \cite{giveon02} using
        Pauldrach SEDs for subsolar, solar and supersolar stellar and
        nebular metallicities. 
        \vspace{1cm}}
        \label{fig:abun}
   \end{SCfigure*}

In this respect, closely related to the problem of determining the
\Teff\ of the ionizing sources is that of their spectral type
classification.
It is illustrative to consider at this point the cases of
G29.96$-$0.02, the Orion nebula,
IRAS\,12073$-$6233, N66 and 30\,Doradus, for which the spectral types
of the ionizing
star are known. The Orion nebula is ionized by the Trapezium
cluster, the dominant source of ionizing radiation being the O6 star
$\theta^1$C\,Ori. 30\,Dor is known to be powered by a stellar cluster
containing a large number of O3--O4 stars \citep{walborn97,massey98}. 
N66 is the most prominent \HII\ region in the SMC. It contains a rich
stellar cluster of at least 33 O stars of which 11 are of type O6.5 or
earlier \citep{massey89}.
Recent high resolution K-band spectral observations of G29 
and IRAS\,12073 \citep[][private communication]{kaper02a,kaper02b} 
allow the determination of the spectral type of the ionizing
stars based mainly upon the presence and strength of photospherical lines
of \ion{C}{iv} and \ion{N}{iii} \citep{hanson96}. These observations
indicate that IRAS\,12073 is ionized by a cluster of young O6--O7.5
stars, while the spectral type of the main
ionizing star in G29 is confirmed to be O3--O5 (we note
that lower resolution K-band spectral observations by
\citealt{watson97b} classified this star as being O5--O8).

As mentioned above, the interpretation of the 
\sratio\ and \neratio\ line ratios observed for these
objects in terms of the stellar \Teff\ depends on the stellar model
used to describe the ionization structure traced by these
ratios.  
The subsequent spectral classification of the ionizing 
source depends on the spectral type attributed to the SED used
in the modelling of the \HII\ region.
We can try to compare the
\Teff\ determined from the observed line ratios by using either CoStar
or Pauldrach SEDS (cf., Fig~\ref{fig:models}) with the observed spectral
type of the ionizing star(s). 
To do so, we need a \Teff--spectral type calibration for O stars.
The \Teff\
calibration of the different spectral types is based on modelling the
observed photospheric hydrogen, helium and metal lines and it is by no
means certain. For instance, the commonly used \Teff--spectral type
calibration by \cite{vacca96} is based on plane-parallel models which do
not incorporate stellar winds and line blocking/blanketing.
The more recent calibration by \cite{martins02}, based on non-LTE
line blanketed atmosphere models with stellar winds computed with the
{\it CMFGEN} code of \cite{hillier98}, gives substantially lower
effective temperatures than the \cite{vacca96} calibration for the
same spectral type. We will use both calibrations.

The result of the comparison between the observed spectral types and the 
predicted stellar properties is summarized in Table~\ref{table:sptype}. 
The calibration by \cite{martins02} gives 
effective temperatures for the observed spectral types $\sim~4000-5000$ K 
lower than \cite{vacca96}.
Even considering this most recent calibration, the CoStar models systematically
predict a lower temperature for the ionizing stars than observed. 
A better match is found by using the Pauldrach SEDs, although the predicted
\Teff\ for G29 is still low. The opposite is true for 
IRAS\,12073, for which the Pauldrach models predict a hotter
star than observed. Moreover,
while our photoionization models
predict that an increase in the degree of ionization is coupled to an
increase in \Teff\ and thus in spectral type, observationally this
trend is less clear (cf., Fig~\ref{fig:sptype}). In particular, G29
and 30\,Dor are both ionized by very early O stars, yet their
ionization ratios are very different. Similarly, IRAS\,12073
and 30\,Dor are ionized by stars with very different  spectral types
but their ionization ratios are very similar. It is clear that the
stellar spectral type as determined from optical or infrared
spectroscopy is not the only controlling factor of the ionization
structure of \HII\ regions. In line with the discussion in
Sect.~\ref{section:ionization} (cf., Fig~\ref{fig:ioniz}), we infer
that the metallicity of the star and \HII\ region is involved as well. 
At this respect, the differences in spectral type between G29 and
  30\,Dor, and in ionization structure between 30\,Dor and IRAS\,12073 
may be a consequence of the differences in metallicity. G29 
has a Ne/H abundance more than 3 times higher than 30\,Dor;
30\,Dor, on the other hand, has a lower abundance than IRAS\,12073
(cf., Fig.~\ref{fig:sptype}).

It is well known that a stellar wind modifies the SED of a star. 
This effect is directly coupled to the strength of
the wind, which in turn depends on the metallicity.    
Metallicity also controls the degree of attenuation of the UV and EUV
stellar radiation due to line blocking and blanketing.
Changes in metallicity can, therefore, transform the stellar
spectrum to that of a cooler or hotter star than what is determined
from the solar metallicity stellar models. 

\section{Influence of the metallicity}
\label{section:metallicity}

The ionization tracer \neratio\ is compared with the metallicity tracer
Ne/H in Fig.~\ref{fig:abun}. The data show a loose trend of
decreasing degree of ionization with metallicity. No selection effect
is thought to influence this relationship. While some of the
LMC/SMC \HII\ regions are among the brightest in these galaxies $-$ 
30\,Dor (LMC) and N66 (SMC) are powered by luminous stellar clusters
containing many early O stars $-$, the LMC sample also contains
sources such as N11A, which are alike to the \HII\ regions in the Galactic
sample in terms of luminosity and Lyman continuum flux. Besides, no
selection effect is apparent in the Galactic \HII\ regions, for which
no obvious variation in the density of the ionized gas or the stellar 
luminosity  with Galactocentric distance is observed (cf., MHP02).
There is some systematic effect, 
however, due to the influence of metallicity on the electron
temperature of the ionized gas. The Galactic Ne/H abundances
were derived using Br$\alpha$ to determine the H$^+$ emission
associated to the nebula and a constant \Tel\ of 7500 K was
assumed. Hence, a high metallicity nebula, which is
expected to have a lower electron temperature, will actually have a
higher Br$\alpha$ emission coefficient than the one adopted in the
evaluation of the Ne/H abundance, and vice versa. 
For reasonable electron temperatures, between 5000 and 10000 K
\citep{shaver83, afflerbach96, deharveng00}, 
this effect is less than a factor of 2
(cf., MHP02). The Magellanic Cloud abundances, however, are not
expected to be influenced by this systematic effect because they were
evaluated using electron temperatures derived from optical
lines. Hence, we consider that the trend in 
Fig.~\ref{fig:abun} is not due to a selection effect, and that there
is a general correlation albeit with a large spread. 
On average, the most highly ionized
\HII\ regions have the lowest metallicity, and vice versa. 

\cite{giveon02} investigated the influence of varying the stellar and nebular
metallicity on the \neratio\ line ratio based upon a set of photoionization
models computed using the Pauldrach SEDs. They found that, for a given
\Teff, the SED hardens with decreasing metallicity due to the
decreased line blocking in the stellar atmosphere
winds. Quantitatively, for a 40 kK star, an increase in the nebular
and stellar metallicity from 0.5Z$_{\sun}$ to 2Z$_{\sun}$ would reduce
\neratio\ by a factor of 20. We compare the results of these
calculations to the observed correlation between metallicity and
degree of ionization (cf., Fig.~\ref{fig:abun}) for 
\Teff=35, 40, 45 kK, Z=0.5Z$_{\sun}$,
1Z$_{\sun}$, 2Z$_{\sun}$ and \nh=800 \cm3.
In view of these results, we see that the large
spread in degree of ionization at a given metallicity observed in our
data reflects a true variation in \Teff. 
Moreover, the observed increase in
the degree of ionization of the \HII\ regions (cf.,
Fig.~\ref{fig:ioniz}) reflects clearly a
hardening of the SEDs due to the decreased metallicity and it
is not necessarily due to a rise of the effective temperature of the 
ionizing star. 

Returning to the individual sources considered in 
Sect.~\ref{section:seds} and in view of Fig.~\ref{fig:abun}, 
the Galactic \HII\ region G29, with a metallicity of roughly 
2Z$_{\sun}$, would be powered
by a star with a temperature slightly higher 
than the one indicated in Fig.~\ref{fig:models}, which are solar metallicity
models. The Magellanic sources N66 and
30\,Dor, with sub-solar metallicities, would be, on the
other hand, ionized by cooler stars because of the decreased line
blocking. However, a proper comparison of the effective temperature derived
from the modelling of the \HII\ region line data with the effective
temperature derived directly from the stellar spectral will require a
proper calibration of stellar spectra for different metallicities. In
particular, the `calibration' of ionizing stars based upon K-band
spectra \citep{hanson96} may require spectra of stars with different
metallicities. Also, the comparison of the strength of hydrogen and
helium lines with predictions of stellar atmospheres 
will require a proper treatment of the influence of metallicity.

\section{The case of starburst galaxies}
\label{section:starburst}

Information on starburst galaxies has been collected in 
order to make a comparison with the \HII\ region results. The data 
obtained for a sample of 9 galaxies observed by ISO are presented in
Table~\ref{table:starburst} and are plotted in Fig.~\ref{fig:abun} as open
stars.  
These galaxies show a similar distribution as
the \HII\ regions: high metallicity--low ionization, low
metallicity--high ionization. In particular, we note that the low metallicity
dwarf irregular galaxy II\,Zw\,40 coincides well with the \HII\ regions
in the LMC, further reinforcing our interpretation that metallicity
and degree of ionization are strongly interconnected. 
Among the low ionized galaxies are M\,82 and NGC\,253, which are the
prototypical starburst galaxies. NGC\,3256, with a \neratio\ similar
to that of M\,82, is a recent galaxy merger. The highly ionized galaxies
include NGC\,5253, a young, Wolf Rayet galaxy, and
NGC\,4038/4039 (the Antennae), where an extensive burst of
star formation is
ongoing in the overlap region between the two merger systems.

\input{H3467T3.tex}

The  \neratio\ line ratio and other ionization ratios observed in 
starburst galaxies have been used to
infer the  spectral type of the ionizing stars and hence, the age of
the starburst
\citep[e.g.][]{schaerer99,crowther99,thornley00,spoon00}. However, as this
discussion has emphasized, the stellar type is only one of the parameters
influencing the ionization structure. 
Furthermore, the uncertain spectral classification of the
ionizing sources through ionization line ratios makes a proper age
estimate of the starburst difficult.
The other factor influencing the ionization structure, metallicity,
measures the cumulative effect of star formation over
the galaxy's history.

\section{Summary and conclusions}
\label{section:summary}

In this paper, we have examined the interplay between the ionization
structure of \HII\ regions, the SEDs of their ionizing stars and
metallicity.
To this end, we combined the near-- and  mid--infrared ISO  spectra 
of a set of Galactic and Magellanic Cloud \HII\ regions. This wavelength range
gives us access to the \SIV\,10.5/\SIII\,18.7 \mum\
and \NeIII\,15.5/\NeII\,12.8 \mum\ line ratios, which are sensitive 
to the ionization structure of the nebula. 
The \sratio\ and \neratio\ line ratios are found
to span a range of more than 2 orders of magnitude and to be
highly correlated. The low metallicity LMC and SMC \HII\ regions overlap 
the trend found for the Galactic objects at the high ionization end.

The dependence of the \sratio\ and  \neratio\ line ratios
on the \Teff\ of the ionizing 
star is investigated using a grid of photoionization models at
solar metallicity. The models are calculated
using two different stellar atmosphere models:
the CoStar SEDs from \cite{schaerer97}  and the more recent SEDs from 
\cite{pauldrach01}. 
The interpretation of the observed ionization line ratios 
in terms of the stellar content of the \HII\ region depends critically 
on the adopted SED.
In particular, we note that the CoStar SEDs predict systematically much
cooler stars than the Pauldrach SEDs for the same ionization conditions.  
As a result, practically the whole observed ionization range can be 
fit by a very narrow  \Teff\ range between $\sim$ 33000 and 42000~K when the
CoStar models are adopted, while the entire range up to 55000~K is 
suitable when using the Pauldrach models. 
These differences originate principally
from the different treatment of the line blanketing/blocking and wind 
properties. 

For a few sources, spectral types of the ionizing sources are presently 
available from infrared (K-band) or optical stellar spectroscopy.
The comparison of these spectral types with those derived indirectly from the 
ionized gas through the \sratio\ and \neratio\ line ratios 
confirms that metallicity has an important 
influence on the stellar spectra.
As impetus for this, we have investigated the relation between the
ionization  structure and  the metallicity comparing the observed
\neratio\ line ratios and  Ne/H abundances. Although a large scatter
is present, a loose correlation between the two parameters is found.
This observed trend is compared to photoionization models calculated
varying  self-consistently the nebular and stellar metallicity
\citep{giveon02}.  The comparison shows that the observed increase in
degree of ionization can be explained by the  hardening of the SED due
to a decrease of the metallicity.

We note, moreover,  that the \Teff\ calibration of the different 
spectral types is based on
modelling the observed photospheric hydrogen, helium and  metal lines. This
calibration is uncertain as it is very sensitive to the input
physics of the stellar model used. The widely used calibration  of
early stars by \cite{vacca96} is based on  plane-parallel, non-LTE
models which do not incorporate stellar winds and line blanketing. It has been
shown, for instance, that the
inclusion of line blanketing, stellar wind and a spherical geometry in the
models \citep{martins02} lowers the
predicted \Teff\ substantially when compared to the calibrated by
\cite{vacca96} for the same spectral type. A new 
calibration is needed if predictions of the stellar content of \HII\ regions
are to be compared with the observed spectral types of the ionizing stars.
Such a recalibration will have to include the effects of metallicity.

In view of this analysis, the main conclusions are the following:

\begin{itemize}

\item[$\bullet$] The interpretation of the observed $X^{+i+1}/X^{+i}$ 
ratios is valid only when the stellar metallicity of the SED 
used in the modelling properly matches the metallicity of the local ISM. 
Not taking into account the effect
of the stellar metallicity correctly can lead to wrong conclusions
on the stellar content of the nebula and thus on the local stellar
population. 

\item[$\bullet$] The adequate treatment of the effect of stellar
wind and line blocking and blanketing on the ionizing spectra of early
O stars, and the availability of grids of stellar models at different
metallicities, is needed.
In this respect, the new models by 
\cite{pauldrach01} constitute an improvement over previous ones.

\item[$\bullet$] A \Teff-spectral type calibration of early stars is
insufficient if the effects of metallicity are not accounted for. A
detailed new calibration is, therefore, necessary.

\item[$\bullet$] The determination of starburst ages in starburst
  galaxies based on $X^{+i+1}/X^{+i}$ ratios is likely incorrect if metallicity
  is not taken into account.

\end{itemize}

Concluding, it is 
important to further study the ionization structure of \HII\ 
regions, particularly using different tracers such as the \HeI\ recombination
lines. Furthermore, from a theoretical point of view, it may be important 
to investigate the origin of the actual spectral classification, such as 
the K band spectral classification via the \ion{C}{iv} and \ion{N}{iii} 
photospherical lines and their dependence on the stellar metallicity.

\begin{acknowledgements}
We thank the anonymous referee, whose comments greatly improved this article.
MICE is supported at MPE by DLR (DARA) under grants 50 QI 86108 
and 50 QI 94023.
\end{acknowledgements}

\end{document}

%% file: H3467T1.tex
\begin{table}[!ht]
\caption{Selected \HII\ regions in the Milky Way, Small Magellanic Cloud
  (SMC) and Large Magellanic Cloud (LMC). Given are the 
  object name, the distance in kpc of the Galactic objects from the
  center of the Galaxy and the coordinates  of the ISO  
  pointing.} 
\label{table:sample}
  \begin{center}
    \leavevmode
    \footnotesize
  \begin{tabular}{lc@{\hspace{1pt}}c@{\hspace{1pt}}c} \hline \\
    \multicolumn{1}{c}{Object} &        
    \multicolumn{1}{c}{R${\rm gal}$} &
    \multicolumn{1}{c}{RA (J2000.0)} &
    \multicolumn{1}{c}{$\delta$ (J2000.0)} \\
    \multicolumn{1}{c}{} &        
    \multicolumn{1}{c}{(kpc)} &
    \multicolumn{1}{c}{(h\ ,\,m\ ,\,s)} &
    \multicolumn{1}{c}{($^o  ,\, \arcmin  ,\, \arcsec$)} \\[5.pt]
\hline \\[1pt]

{\bf Milky Way:}\\[5pt]
IR\,01045$+$6506 & 13.8 & 01 07 50.7 & $+$65 21 21.7\\
IR\,02219$+$6125 & 11.0 & 02 25 44.6 & $+$62 06 11.3\\
IR\,10589$-$6034 &  9.5 & 11 00 59.8 & $-$60 50 27.1\\
IR\,11143$-$6113 &  9.7 & 11 16 33.8 & $-$61 29 59.4\\
IR\,12063$-$6259 &  9.3 & 12 09 01.1 & $-$63 15 54.7\\
IR\,12073$-$6233 & 10.1 & 12 10 00.3 & $-$62 49 56.5\\
IR\,12331$-$6134 &  6.9 & 12 36 01.9 & $-$61 51 03.9\\
IR\,15384$-$5348 &  6.4 & 15 42 17.1 & $-$53 58 31.5\\
IR\,15502$-$5302 &  4.6 & 15 54 06.0 & $-$53 11 36.4\\
IR\,17221$-$3619 &  5.2 & 17 25 31.7 & $-$36 21 53.5\\
IR\,17455$-$2800 &  0.5 & 17 48 41.5 & $-$28 01 38.3\\
IR\,18116$-$1646 &  4.3 & 18 14 35.2 & $-$16 45 20.6\\
IR\,18317$-$0757 &  4.5 & 18 34 24.9 & $-$07 54 47.9\\
IR\,18434$-$0242 &  4.6 & 18 46 04.0 & $-$02 39 20.5\\
IR\,18479$-$0005 &  7.5 & 18 50 30.8 & $-$00 01 59.4\\
IR\,18502$+$0051 &  4.7 & 18 52 50.2 & $+$00 55 27.6\\
IR\,19598$+$3324 &  9.8 & 20 01 45.6 & $+$33 32 43.7\\
Dr\,21           &  8.6 & 20 39 00.9 & $+$42 19 41.9\\
IR\,21190$+$5140 & 12.7 & 21 20 44.9 & $+$51 53 26.5\\
IR\,23030$+$5958 & 11.4 & 23 05 10.6 & $+$60 14 40.6\\[0.1cm]
\hline \\[1pt]

{\bf SMC:}\\[5pt]
N66              &  & 00 59 03.7 & $-$72 10 39.9\\
N81              &  & 01 09 13.6 & $-$73 11 41.1\\[0.1cm] 
\hline \\[1pt]
{\bf LMC:}\\[5pt]
N4A              & & 04 52 08.4  & $-$66 55 23.4\\
N83B             & & 04 54 25.2  & $-$69 10 59.8\\
N11A             & & 04 57 16.2  & $-$66 23 18.3\\
30\,Dor\#1       & & 05 38 33.5  & $-$69 06 27.1\\
30\,Dor\#2       & & 05 38 35.5  & $-$69 05 41.2\\
30\,Dor\#3       & & 05 38 46.0  & $-$69 05 07.9\\
30\,Dor\#4       & & 05 38 54.2  & $-$69 05 15.3\\
N160A1           & & 05 39 43.3  & $-$69 38 51.4\\
N160A2           & & 05 39 46.1  & $-$69 38 36.6\\
N159-5           & & 05 40 02.4  & $-$69 44 33.4\\[0.1cm]
\hline 
  \end{tabular}
  \end{center}
\end{table}

%%% Local Variables: 
%%% mode: latex
%%% TeX-master: t
%%% TeX-master: t
%%% TeX-master: t
%%% TeX-master: t
%%% TeX-master: t
%%% TeX-master: t
%%% TeX-master: t
%%% End: 

%% file: H3467T2.tex
\begin{table*}[!ht]
  \caption{Comparison of the stellar properties predicted via the 
  \neratio\ and \sratio\ line ratios with the spectral type derived
  directly from infrared/optical stellar spectra.
  Given are the object name, the observed spectral type of the ionizing star(s)
  with its estimated \Teff, the predicted stellar properties
  by using either the CoStar or the Pauldrach SEDs and the reference
  of the observed spectral type. The
  objects are sorted by increasing degree of ionization.}
  \label{table:sptype}
  \begin{center}
    \leavevmode
    \normalsize
    \begin{tabular}[h]{llcccccc}
      \hline \\[-7pt]
    \multicolumn{1}{c}{} &
    \multicolumn{1}{c}{} & 
    \multicolumn{2}{c}{Observed \Teff$^\dagger$} &
    \multicolumn{1}{c}{} &
    \multicolumn{2}{c}{Predicted \Teff$^\dagger$} &
    \multicolumn{1}{c}{}\\ \cline{3-4} \cline{6-7} \\[-5pt]
    \multicolumn{1}{c}{Object name} &
    \multicolumn{1}{c}{Spectral Type} & 
    \multicolumn{1}{c}{($a$)} &
    \multicolumn{1}{c}{($b$)} &
    \multicolumn{1}{c}{} &
    \multicolumn{1}{c}{CoStar} &
    \multicolumn{1}{c}{Pauldrach} &
    \multicolumn{1}{c}{Ref.}\\[5pt] \hline \\[-7pt]
    G\,29.96$-$0.02    & O5--O3           & 46--51 & 42--49 && ~33--36$^\star$ &35--40 & 1\\
    Orion nebula       & O6               & 44     & 39     && 33--36 &40--45 & 2\\
    N66                & O6.5 and earlier & $>42$  & $>37$  && 33--42 &$>$40  & 3\\
    30\,Doradus        & O4--O3           & 49--51 & 43--49 && 36--42 &$>$45  &4,5\\
    IRAS\,12073$-$6233 & O7.5--O6         & 40--44 & 36--39 && $>36$  &$>$45  &1\\
    \hline \\[-5pt]
    \end{tabular}
  \end{center}
($a$) \Teff\ derived from the observed spectral type using the calibration by \cite{vacca96}; 
($b$) \Teff\ derived from the observed spectral type using the calibration by \cite{martins02};
($\dagger$) \Teff\ in kK;
($\star$) The detailed photoionization model by
\cite{morisset:paperiii}, which uses the CoStar SEDs, predicts a
\Teff\ of  $\sim 30^{+2}_{-1}$ kK.
REFERENCES: (1) \cite{kaper02a,kaper02b} and private communication; (2) SIMBAD
database; (3) \cite{massey89}; (4) \cite{walborn97}; (5) \cite{massey98}.
\end{table*}

%%% Local Variables: 
%%% mode: latex
%%% TeX-master: t
%%% TeX-master: t
%%% End: 

%% file: H3467T3.tex
\begin{table}[!ht]
  \caption{\neratio\ line ratios and Ne/H abundances for a sample of
    galaxies, including Sgr\,A$^*$. The sample is
  sorted out by increasing degree of ionization.}
  \label{table:starburst}
  \begin{center}
    \leavevmode
    \normalsize
    \begin{tabular}[h]{lccc}
      \hline \\[-7pt]
    \multicolumn{1}{c}{Object} &
    \multicolumn{1}{c}{\neratio} &
    \multicolumn{1}{c}{Ne/H$^\dagger$} & 
    \multicolumn{1}{c}{Ref.} \\
    \multicolumn{1}{c}{name} &
    \multicolumn{1}{c}{} &
    \multicolumn{1}{c}{($10^{-4}$)} & 
    \multicolumn{1}{c}{} \\[2pt] 
\hline \\[-7pt]
M83            & 0.05 &   2.1 & 1,$+$\\
Sgr\,A$^*$     & 0.06 &   1.8 & 2,$+$\\
NGC\,7552      & 0.08 &   1.8 & 1,$+$\\
NGC\,253       & 0.08 &   1.8 & 1,$+$\\
NGC\,4945$^\ddagger$      & 0.09 &   1.3 & 3    \\
M82            & 0.18 &   1.4 & 4    \\
NGC\,3256      & 0.20 &   2.3 & 5    \\
NGC\,4038/39$^\diamond$   & 0.92 &   0.6 & 6    \\
NGC\,5253      & 4.0  &   0.7 & 7    \\
II\,ZW\,40     &   12 &   0.4 & 1,$+$\\[2pt]
      \hline \\
    \end{tabular}
  \end{center}
($\dagger$) Calculated from the mid-infrared \NeII~12.8~\mum,
\NeIII~15.5~\mum\ and Br$\alpha$ 4.0~\mum\ lines assuming \Tel$=10^4$ K and the
low density limit. No extinction correction is applied.
($\ddagger$) The nature of this galaxy is uncertain. The infrared
diagnostics show no evidence for the existence of the AGN inferred
from hard X-ray observations \citep[see][]{spoon00}.
($\diamond$) ISO pointing centered on the overlap region between the two
interactinc systems.\\
REFERENCES:  (1) \cite{thornley00}; (2) \cite{lutz96};
(3) \cite{spoon00}; (4) \cite{forster01}; 
(5) \cite{rigopoulou96}; 
(6) \cite{kunze96}; (7) \cite{crowther99};
(+) ISO archive.
\end{table}

%%% Local Variables: 
%%% mode: latex
%%% TeX-master: t
%%% TeX-master: "\\"
%%% TeX-master: t
%%% TeX-master: t
%%% TeX-master: t
%%% TeX-master: t
%%% End: 